\documentclass[sn-mathphys-num]{sn-jnl}

\usepackage{graphicx}%
\usepackage{multirow}%
\usepackage{amsmath,amssymb,amsfonts}%
\usepackage{amsthm}%
\usepackage{mathrsfs}%
\usepackage[title]{appendix}%
\usepackage[dvipsnames]{xcolor}
\usepackage{textcomp}%
\usepackage{manyfoot}%
\usepackage{booktabs}%
\usepackage{algorithm}%
\usepackage{algorithmicx}%
\usepackage{algpseudocode}%
\usepackage{listings}%

\theoremstyle{thmstyleone}%
\theoremstyle{thmstyletwo}%

\theoremstyle{thmstylethree}%

\begin{document}

\title[]{The enigmatic gravitational partition function}


\author*[1]{\fnm{Batoul} \sur{Banihashemi}}\email{bbanihas@ucsc.edu}
\author*[2]{\fnm{Ted} \sur{Jacobson}}\email{jacobson@umd.edu}



\affil[1]{\orgdiv{Physics Department}, \orgname{UC Santa Cruz}, \orgaddress{\street{1156 High Street}, 
\city{Santa Cruz}, \postcode{95064}, \state{California}, \country{USA}}}

\affil[2]{\orgdiv{Maryland Center for Fundamental Physics}, \orgname{University of Maryland}, \city{College Park}, \postcode{20742}, \state{Maryland}, \country{USA}}



\abstract{The saddle point approximation to 
formal quantum gravitational partition functions
has yielded plausible computations of horizon entropy
in various settings, but  it stands on shaky ground. 
In this paper we visit some of that shaky ground,
address some foundational questions, 
and describe efforts toward a more solid footing.
We focus on the case of
de Sitter horizon entropy which, it has been argued, corresponds
to the dimension of the Hilbert space of a ball of 
space surrounded by the cosmological horizon.
}

\keywords{quantum gravity}



\maketitle

\section{Introduction}\label{sec1}

Nearly a century has passed since the discovery of quantum theory, yet we still lack a fundamental understanding of how it fits together with general relativity. Although we have a sensible approximation scheme, the foundations remain deeply elusive for several reasons. One of our few precious clues is black hole entropy, and more generally the entropy of causal horizons. But we understand that entropy today only through a glass darkly. The quantum gravitational partition function, introduced nearly half a century ago, appears to miraculously capture that entropy, as well as other elements of quantum gravity truth, but it stands on shaky ground. In this paper we visit some of that shaky ground and describe some efforts toward a more solid footing.  

The paper is based in part on a talk given by TJ to a broad audience at the Lema\^{i}tre Conference, 2024.
It focuses on the quantum gravitational significance of the
entropy of black hole and de Sitter horizons. 
It is fitting therefore to recall at the outset that 
Lema\^{i}tre was the first to understand the occurrence of horizons in general relativity.
In 1925 he
explained the physical nature of the coordinate singularity in the static de Sitter metric \cite{lemaitre1925note}, in 1927 he used the word ``horizon'' to describe it \cite{Lemaitre:1927zz,2013GReGr..45.1635L}, and in 1933, he was the first to understand the Schwarzschild ``singularity'', stating \cite{1933ASSB...53...51L,1997GReGr..29..641L}\\

\vspace{0mm}
\begin{minipage}{11.5cm}
    \textit{The singularity of the Schwarzschild field is thus a fictitious singularity, analogous to that which appears at the horizon of the centre in the original form of the de Sitter universe.}
\end{minipage}

\vspace{3.5mm}
\noindent He thus not only explained the nature of the horizon, but also gave it its name, 
twenty-five years before the famous Finkelstein paper \cite{Finkelstein:1958zz} with the one-way membrane picture.

\section{Partition function}\label{sec2}
In their seminal 1977 paper,  
\textit{Action integrals and partition functions in quantum gravity} \cite{Gibbons:1976ue},
Gibbons and Hawking (GH)
had the audacious idea that
one could extract thermodynamic and statistical insights
about quantum gravity 
from a path integral 
for the partition function, formally emulating what 
is standard procedure for a nongravitational quantum system.
Such a system has a thermal partition function in the canonical
ensemble given by
\begin{equation}\label{Z}
    Z = {\rm Tr} e^{-\beta H}
\end{equation}
where $H$ is the Hamiltonian and $\beta$ is the inverse temperature.
From $Z$ one obtains thermodynamic quantities, e.g.,  
free energy, internal energy, and entropy.

For a quantum system 
one can view $e^{-\beta H}$ as the 
time evolution operator 
for an imaginary time interval 
$\Delta t = -i\hbar\beta =: -i \Delta \tau$, and 
$Z$ can thus be expressed as a path integral
over trajectories 
that are periodic in imaginary time \cite{feynman2010quantum}. 
For a particle in a potential $V(x)$ this takes the form
\begin{equation}
    Z =\int_{x\;{\rm period}\; \hbar\beta} \!\!\!\!\!\!\!\!\!\!\!\!\!\!\!\!\!\!\!{\cal D}x\; e^{-I_{\rm E}/\hbar},
\end{equation}
where $I_{\rm E}$ is the ``Euclidean action'', 
\begin{equation}
    I_{\rm E} = \int d\tau \;\left[\frac12 \left(\!\frac{dx}{d\tau}\!\right)^{\!\!2} + V(x)\right].
\end{equation}
In quantum field theory, one writes a similar expression, with 
the particle path $x(\tau)$ replaced by a path in the space of field configurations, $\phi(\vec x,\tau)$.

In general relativity, especially in the presence of horizons,
things are not so clear, for several
reasons: 
\begin{quote}
\begin{itemize}
    \item $H$ is a boundary term
    \item initial value constraints must be imposed
    \item the Euclidean action is unbounded below
    \item “time” depends on the metric and the loop it is measured on
    \item time flow is null on a horizon
    \item there is a spacelike singularity inside a black hole
\end{itemize}
\end{quote}
Without pretending to resolve fundamental issues, Gibbons and Hawking 
focused on candidate Euclidean signature saddle points of the
path integral, which close off smoothly 
at the Euclidean horizon,
so that the black hole interior plays no role.
The saddle point action yields the free energy, 
and from that they recovered the 
Bekenstein-Hawking entropy $A/4\hbar G$ 
for black holes and de Sitter space,
which had previously been found by combining the classical 
laws of horizon mechanics with Hawking's horizon temperature formula
obtained from quantum field theory in curved spacetime.
This agreement suggests that perhaps more about quantum gravity can be learned by better understanding some questions about this partition function:\\  

%
\begin{minipage}{14cm}
\noindent \hspace{0cm} \textit{What integral is approximated by this saddle point?}\\

\noindent \hspace{0cm} \textit{What is the integration contour, and why is different topology allowed?}\\
\end{minipage}

\noindent We will first review how the Gibbons-Hawking analysis works for (nonrotating) black holes and for de Sitter horizons, and will then discuss attempts to better understand the 
foundations of the gravitational partition function 
from first principles. 

\section{Gibbons-Hawking partition function}\label{sec3}

Gibbons and Hawking considered the calculation of
\eqref{Z} for asymptotically flat 
black hole spacetimes and for spatially closed 
de Sitter spacetimes.\footnote{If the
boundary is at asymptotically flat infinity the canonical 
ensemble is ill defined.
This problem can be avoided by considering asymptotically Anti-de Sitter spacetimes \cite{HawkingPage}, or finite boundaries \cite{York86}.
For brevity here we assume this issue has been suitably dealt with.} 
To begin, we discuss asymptotically flat spacetimes,
and restrict to a sector where the bounded region
contains no horizon. 
For example, we might be interested in the 
quantum statistical mechanics of a star, 
under conditions where black hole states are not relevant. 
Next we'll discuss the case where a black hole horizon is 
present, and finally we'll turn to the spatially closed case
without boundary.

\subsection{No horizon}

\begin{figure}[h]
\centering
\includegraphics[width=0.3\textwidth]{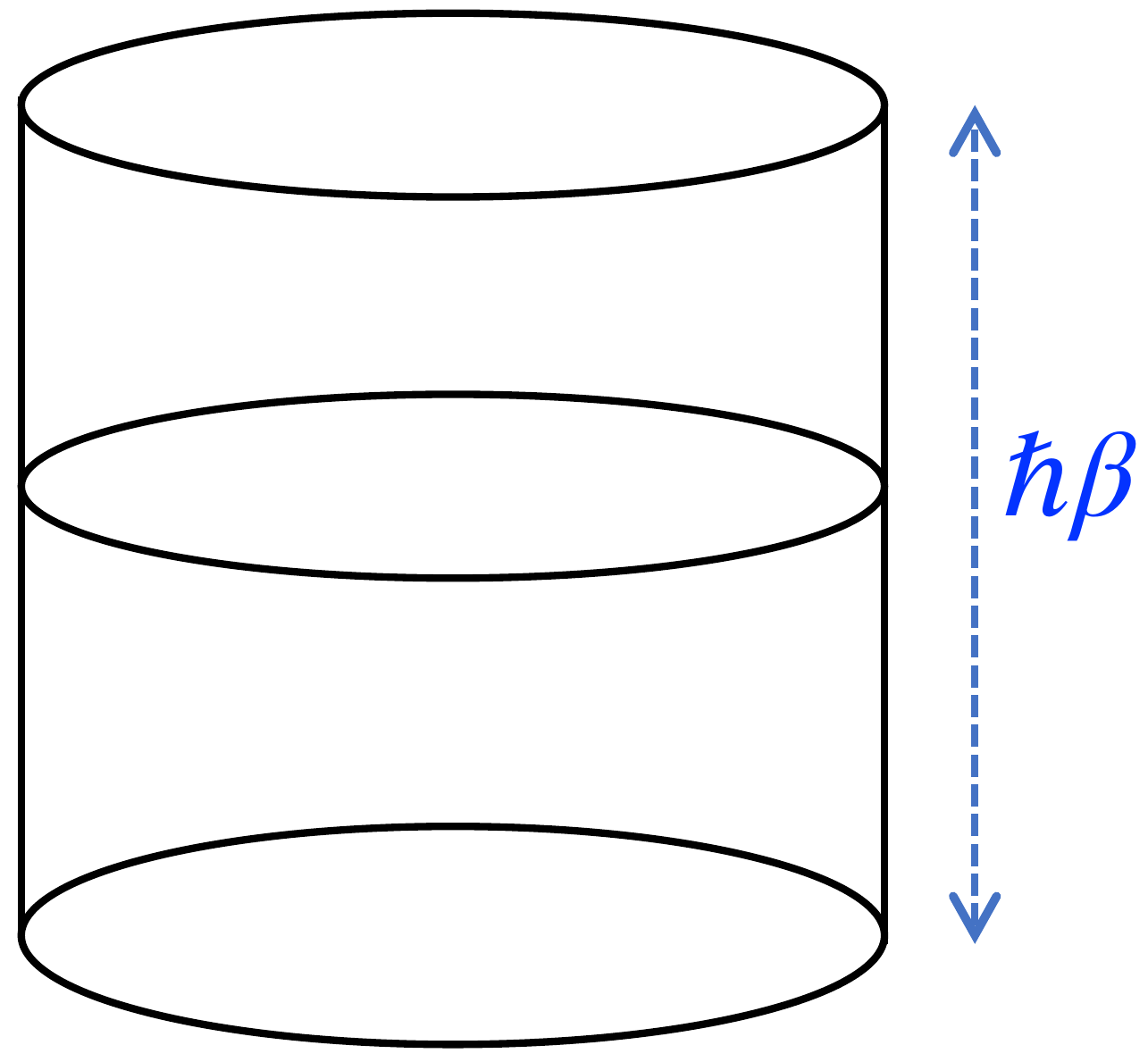}
\caption{Topology of the paths  in the path integral for 
the partition function
$Z$ \eqref{Z} for the canonical ensemble of a gravitating system 
with an outer system boundary, and no inner boundary. 
The top and bottom of the cylinder are identified, 
and $\hbar \beta$ is the time period of the Euclidean metric 
fixed at the boundary.}\label{soupcan}
\end{figure}

Each time slice of the path integral has the topology
of a spatial ball whose boundary is the system boundary.
Since the path integral is computing a trace,
the imaginary time dimension is periodic.  In Fig.\ \ref{soupcan} 
this is illustrated with two-dimensional space, 
so the spatial ball is a 2-disc, but with three spatial dimensions
it would be a 3-ball. The top and bottom of the
cylinder are identified, so the
path topology is $B^3\times S^1$.
In order to specify a thermal ensemble,
stationary boundary conditions are imposed at the ball boundary,
and a time foliation and boundary time flow 
are specified. The Hamiltonian, which in a generally covariant theory 
is a boundary term, generates the corresponding time evolution.
For the canonical ensemble, $\beta$ 
is identified with the proper Euclidean boundary time flow period.

Since the evolution time is imaginary 
it is natural to think, if the path integral
can be expressed in a covariant rather than canonical form,
that the paths for the gravitational degrees of freedom 
should be Euclidean spacetime geometries.
But this is not correct. That it cannot 
be correct is evident from the fact that 
the Euclidean Einstein-Hilbert action is unbounded below \cite{GHP}, 
so the integral would diverge. To diagnose the etiology of this disease
one should start from first principles. 
A path integral representation of the trace in $Z$ should be a sum over states in Hilbert space, so one must restrict to paths that label sequences of states in Hilbert space. 
This means that one must impose the diffeomorphism 
constraints of the theory, which restrict initial
data to live in a physical phase space, 
and one must fix the gauge (diffeomorphism) freedom 
or somehow divide by the volume of the gauge group. 
The imposition of the constraints eliminates those configurations 
with rapidly varying conformal factor that make 
the action unbounded below, and with asymptotically flat
boundary conditions it ensures that at least the linearized action is 
bounded below \cite{Hajicek, Schleich, Hartle-Schleich, Mazur-Mottola, Loll-Dasgupta}.
The broader point, however, is that the issue of the negative
unboundedness of the Euclidean action goes away if one
starts from the reduced phase space path integral.

A standard method of imposing the constraints
is to start with a phase space path integral, include Dirac delta functions
of the constraints, and express those delta functions as Fourier 
integrals over {\it real} ``Lagrange multipliers''. 
After the conjugate momenta have been integrated out,
these Lagrange multipliers wind up playing the
role of lapse and shift in the {\it Lorentzian} 
spacetime metric appearing in the action
 \cite{Faddeev:1973zb}. 
But this entails a mismatch with the boundary lapse 
(let's suppose the boundary shift vanishes), 
which must be imaginary since we are computing trace of the 
imaginary time evolution operator $e^{-\beta H}$.
The Lagrange multiplier contours can be deformed so as to meet 
an imaginary boundary condition,
but they cannot be pure imaginary in the bulk 
since that would take us back to the Euclidean action 
for which the constraints are not imposed by the path integral.
Therefore the path integral contour must be over complex metrics,
if it's over spacetime metrics at all. 
If the Gibbons-Hawking saddle is indeed relevant,
it must be reached by contour deformation
from the defining contour. For references and 
a detailed discussion of these 
issues,  see \cite{Banihashemi:2022jys}. 

\subsection{Black hole}

\begin{figure}[h]
\centering
\includegraphics[width=0.5\textwidth]{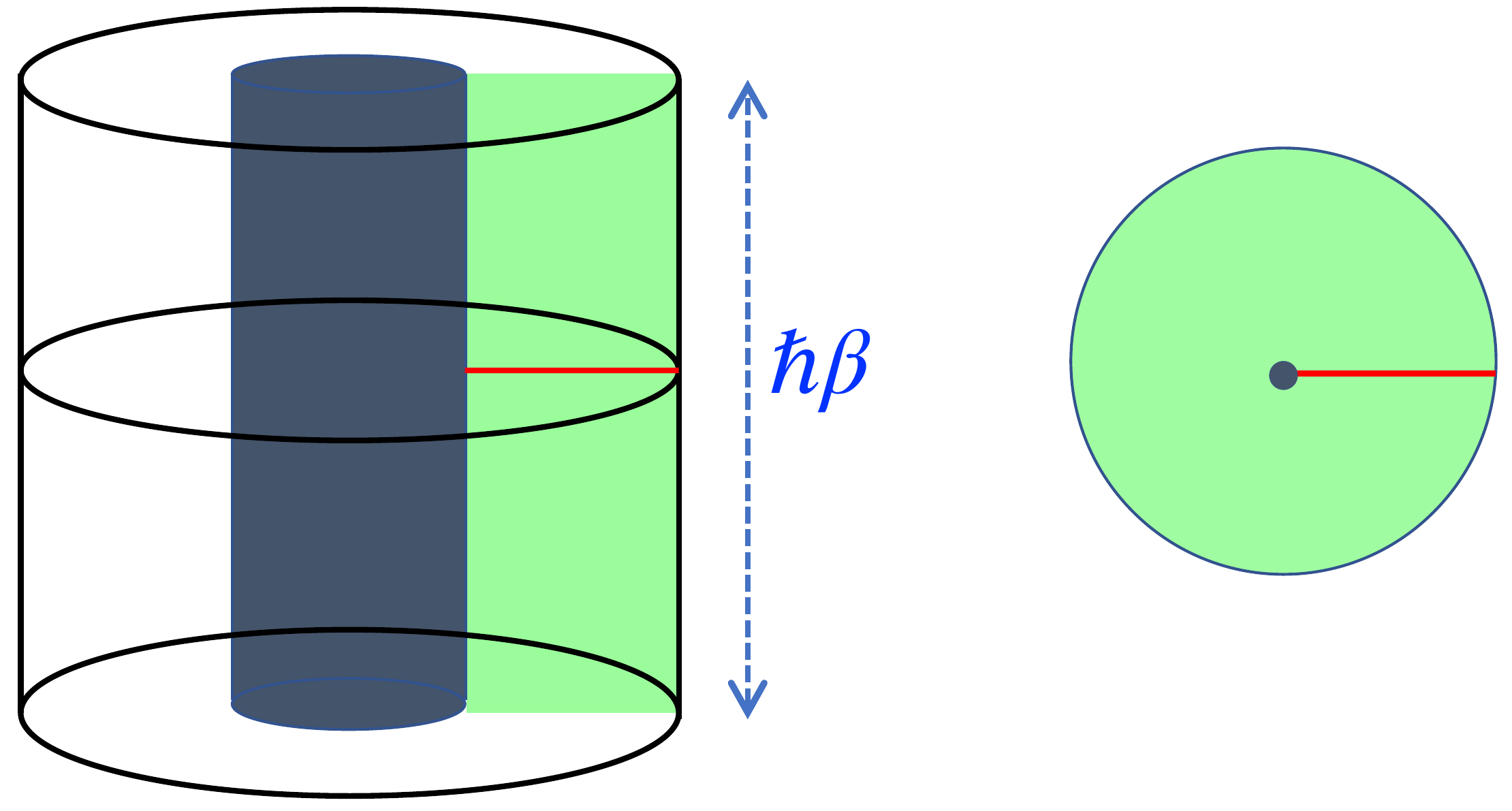}
\caption{Topology of the paths for 
the partition function
$Z$ \eqref{Z} for the canonical ensemble of a gravitating system 
with an outer system boundary
like in Fig.~\ref{soupcan},
but containing also an inner boundary at a black hole horizon. The black cylinder is
the world tube of the horizon. The lapse---and thus the time flow---is 
taken to vanish at the horizon, so as not to impose constraints there and not 
to produce an evolving inner boundary of the system. At the Euclidean saddle the vertical 
extension 
along the black cylinder therefore vanishes,
hence 
the black cylinder is squashed to a circle (a 2-sphere if the spacetime is 4-dimensional). The periodically
identified green strip between the outer boundary and the horizon has the topology 
of a disc, as shown on the right, where the black dot in the center lies on the horizon,
and the perimeter is an outer boundary loop of circumference $\hbar \beta$.
}\label{bhsoup}
\end{figure}

Now let's add a black hole to the state sum (see Fig.~\ref{bhsoup}).
In addition to the outer boundary, there's an inner boundary on each spatial
slice, at the horizon. The region inside the horizon is the solid grey cylinder
in the figure. If this were a Lorentzian spacetime diagram of a Schwarzschild
black hole for example, the vertical direction on the 
surface of the grey cylinder would be null, the horizontal spatial 
slices could be surfaces of constant Painlev\'{e}-Gullstrand time coordinate,
and the axis of the grey cylinder would be the spacelike singularity at $r=0$.
One could try to make sense of the path integral over geometries like
this, but as far as we know that has not been attempted. Moreover, it
would be topologically disconnected from the GH saddle.

The GH saddle is the Euclidean Schwarzschild geometry.
The red radial line in Fig.~\ref{bhsoup}
extends from the outer boundary to a point on the horizon.
If this red line is translated in
Euclidean time in the GH saddle geometry, 
and the initial and final lines 
are identified, it sweeps out a disc, because a
vertical null curve along the horizon, being a Euclidean 
zero length curve, is actually a single point. This disc is
illustrated on the right in Fig.~\ref{bhsoup}.
The full topology of the saddle is the product of this 
disc with the 2-sphere of angular dimensions,
$S^2\times D^2$. A horizontal slice of this Euclidean
geometry coincides with a static spatial slice of the 
Lorentzian Schwarzschild solution terminating
at the bifurcation surface of the horizon. So the
GH saddle geometry corresponds to only the part of the
spacetime outside horizon. It is therefore natural 
to interpret the entropy that results from this 
approximation to $Z$ as the entropy
of the degrees of freedom {\it outside the horizon}. 

The entropy is computed from the saddle point
approximation to the partition function as
\begin{equation}
    S = (1-\beta\partial_\beta)\ln Z \approx (\beta\partial_\beta -1)I^{\rm saddle}_{E}/\hbar = A/4\hbar G,
\end{equation}
where $I^{\rm saddle}_{E}$ is the Euclidean saddle action and $A$ is the horizon area of the saddle. The saddle is the Euclidean solution to the Einstein equation that meets the outer
boundary condition, where the spatial geometry and proper time period $\beta$ 
define the ensemble, and that is smooth (not conically singular) 
at the Euclidean horizon. 
For a spherically symmetric Euclidean vacuum solution, 
in suitable coordinates
the line element takes the form
\begin{equation}\label{ES0}
    ds^2 = [N(r)]^2 d\tau^2 + [N(r)]^{-2}dr^2 + r^2 d\Omega^2,
\end{equation}
where $N({\rm horizon})=0$. In terms of the coordinate $l$ that measures
proper length on the constant $\tau$ surfaces and vanishes 
at the horizon, 
we have
\begin{equation}\label{ES}
    ds^2 = (\kappa l + \dots)^2 d\bar\tau^2 + dl^2 + r^2 d\Omega^2,
\end{equation}
where $\kappa$ is the surface gravity defined with respect to the Killing vector
$\partial_{\bar\tau}$, and we have adopted the rescaled Euclidean 
time coordinate $\bar\tau$ that
agrees with proper time at the boundary.
The temperature of the ensemble determines the range of $\bar\tau$ to be  
$\Delta\bar\tau = \hbar \beta$.
Smoothness at the horizon requires that as $l\rightarrow0$ the $l$-$\bar\tau$
line element in \eqref{ES} agrees with that of the flat plane in polar coordinates,
which requires that $\kappa\Delta\bar\tau= 2\pi$. It follows that 
$\kappa = 2\pi/\hbar\beta$, which determines $\kappa$, and therefore
the mass of the Schwarzschild solution, as a function of $\beta$.
This implies that the boundary temperature $T_B$ is equal to  
the Hawking temperature for the saddle black hole, 
$T_B = \hbar\kappa/2\pi = T_H$.

\subsection{de Sitter horizon}

Gibbons and Hawking also applied their formalism to the case of 
de Sitter space (dS), a solution to the Einstein equation
with a positive cosmological constant
$\Lambda$. 
They argued in \cite{Gibbons:1977mu}
that laws similar to the laws of black hole mechanics apply to a 
static patch of de Sitter space, and that the  area $A$ of the horizon of a static
patch is associated with a Bekenstein-Hawking entropy $A/4G\hbar$.
To relate this to a partition function in \cite{Gibbons:1976ue} they
reasoned that, since there is no outer boundary, 
the Hamiltonian and therefore the 
internal energy vanishes, so the free energy $F = E - TS$
reduces to $-TS$, hence the partition function reduces to 
$Z \approx e^S$, which identifies the saddle point approximation to the 
entropy as $S = -I^{\rm saddle}_{E}/\hbar$.  They took the saddle to be
the round 4-sphere, which is a Euclidean solution to the Einstein equation
with the cosmological constant $\Lambda$, and is the solution with the lowest action.
Remarkably, this yields for $S$ the Bekenstein-Hawking entropy.

Not discussed in \cite{Gibbons:1976ue}, however, is what ensemble this is the entropy \textit{of},
although the answer to that question can be read from the formalism and
what has been said so far. To begin with, if the Hamiltonian vanishes, 
then the partition function \eqref{Z} becomes 
\begin{equation}\label{TrI}
    Z = {\rm Tr}\, I,
\end{equation}
where $I$ is the identity operator on the Hilbert space. That is, $Z$ is just the dimension
of the Hilbert space. But what is the system whose Hilbert space $Z$ is counting the dimension of? 

\begin{figure}[h!]
\centering
\includegraphics[width=0.3\textwidth]{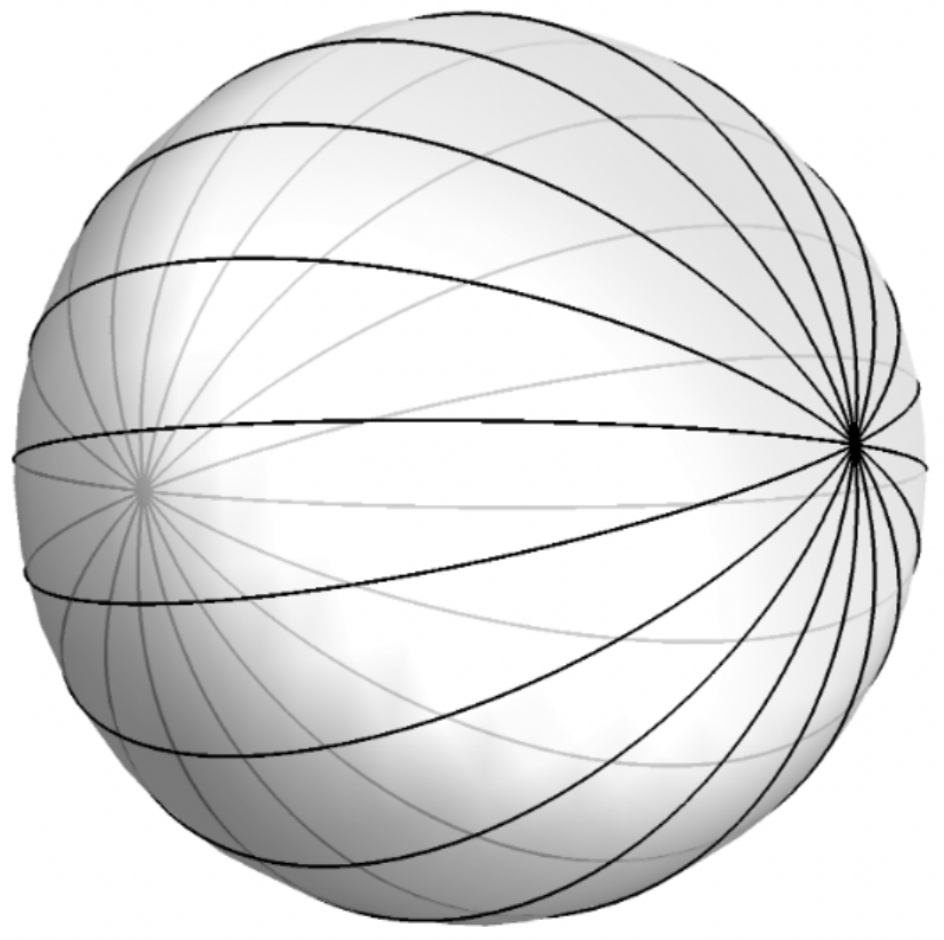}
\caption{Two-dimensional Euclidean dS. A spatial slice of the static patch is a 1-ball, which is a line segment, and its ``surface'' is a pair of points comprising the horizon. Evolving a line segment through a Euclidean time circle with vanishing circumference at the endpoints yields a 2-sphere. \label{2sphere}}
\end{figure}

In the case of the black hole we argued that the system consists of the degrees of freedom outside the horizon, i.e. on the side of the horizon where the system boundary lies. 
This was based on the fact that the saddle whose action yields the entropy
is the evolution of a spatial slice outside the horizon, through a Euclidean time circle with 
vanishing circumference at the horizon. 
The dS case is like an ``inside-out'' black hole. The horizon surrounds the observer 
who is located in the center of the static patch, and there is no system boundary where that observer is located. 
A spatial slice of the static patch is a 3-ball whose surface is the horizon. 
And evolution of a 3-ball through 
a Euclidean time circle with vanishing circumference at the surface of the ball yields a 4-sphere $S^4$, which is the topology of the GH dS saddle. 
(To see why a 4-sphere results, it helps to consider the situation in two
fewer dimensions, depicted in Fig.~\ref{2sphere}.)

The upshot is that the system in this case consists of the degrees of freedom
of a ball of space; and the Bekenstein-Hawking entropy “of de Sitter space” 
is actually the log of the dimension of the entire Hilbert space 
of states of a ball of space, for a fixed value of the cosmological constant. 


\section{Seeking a solid foundation}

The physical interpretation of the GH gravitational 
partition function we have been discussing is based on formal and somewhat
fuzzy reasoning. And it relies heavily on properties of the saddle points,
rather than on fundamental properties of the partition function.
It is taken seriously largely because it yields an entropy that is consistent with
the semiclassical entropy that emerges from sharper, well founded 
considerations (namely, classical horizon mechanics 
together with the Hawking temperature computed using quantum field theory in curved spacetime \cite{Bekenstein:1973ur, Hawking:1975vcx, Jacobson:2003wv},
string theory computations of the entropy of special black holes in supergravity
theories, and the AdS/CFT duality that emerged from the 
latter \cite{DeHaro:2019gno}).
With the aim of probing for a more solid foundation, 
let us pose some key questions that should be answered about 
these gravitational partition functions.
In the following we attempt to address these questions. Our answers 
are at present the best we can offer. We make no claims of certainty!

\paragraph{What microstates are counted by the entropy?} 
As we have suggested above, the states are those accessible 
to an observer outside the horizon. These include the vacuum 
fluctuations that are entangled with partners behind the horizon. 
And in the case of gauge and gravitational fields, 
they include the “edge modes”, 
i.e., the “would-be gauge degrees of freedom'' \cite{Carlip:1994gy, Balachandran:1994up, Carlip:1995cd} 
that are rendered physical by virtue of the unobservability 
of the region behind the horizon.

\paragraph{How does $Z$ “know” the microstates without actually counting them?}
It knows the number of microstates because that is encoded in the gravitational 
constant $G$.
$G$ is a measured, 
macroscopic  parameter of the low energy effective theory, whose value
is determined by the microscopic  physics. 
Entropy for a material system may similarly depend on macroscopic
parameters, like elastic moduli, but in that case it 
also depends on microscopic parameters
like number density and molecular structure, as well as on temperature.
For horizon entropy, in contrast, only
the bending modulus of spacetime 
(i.e., the strength of the gravitational interaction) is needed
to determine its area-density, $(4G\hbar)^{-1}$. 
There is no temperature dependence, simply because
horizon entropy is a property of the vacuum, 
but the lack of dependence on any microscopic parameters 
is striking and demands explanation.
The universal relation between
the bending modulus of spacetime and horizon entropy
density
points to a deep equivalence  
between these properties.
In fact,
one can infer the existence of gravity and the form of the 
Einstein field equation from the assumption that 
the area density of vacuum entropy is finite 
and universal \cite{Jacobson:1995ab,Jacobson:2012yt}.

\paragraph{Can it be trusted, given the UV incompleteness of GR?}
Yes, since the relevant summary of the microphysics
is captured by the value of $G$, which appears in 
the low energy effective action of the saddle that 
yields the leading approximation to the entropy.
Unlike what happens for nongravitational theories,
the entropy is captured by a classical saddle action
because that action is enormous compared 
to Planck's constant. Whether a similar thing could happen 
with another physical system is unknown to us.

\paragraph{Which topologies should be included?}
For a nongravitational field theory (as explained above) there is no
question: the paths in the path integral
representation of the partition
function are field configurations on the spatial domain $\Sigma$ that are
periodic in (imaginary) time, so the topology of the manifold on which the fields
live is $\Sigma\times S^1$. The field configurations at each time label
a basis for the configuration eigenstates that span the Hilbert space.

Unless one departs from the rules of quantum mechanics, the same thing
should presumably be true in the gravitational case, 
although the rules start to become murky when ``the system'' is 
a subregion defined by dynamical conditions, like the region outside a
black hole horizon. Gibbons and Hawking sidestepped the murkiness by
focusing on the Euclidean saddle that was plausibly relevant to the
computation. That saddle would be of the form 
$\Sigma\times S^1$, where $\Sigma$ is the 
spatial region outside the horizon, but for the 
topological transmutation brought about by the vanishing 
of the $S^1$ circumference at the Euclidean horizon. 
That transmutation, needed for smoothness of the 
Euclidean saddle without boundary, was understood
long ago to be a geometric necessity in order for the
fluctuations around the saddle to correspond to the
analytic continuation of ``normal'' Lorentzian vacuum
fluctuations \cite{Gibbons:1976es,Gibbons:1976pt}. 

The need to impose this geometric smoothness condition can 
be appreciated by pointing out an overly formal statement
made above. Namely, the configuration eigenstates of a
quantum field are not really in the physical Hilbert space.
The Schr\"{o}dinger picture Hilbert space for a quantum field theory
is not all possible functionals of the field configuration.
Rather, it is only those functionals that look like the Minkowski
vacuum at short distances. The smoothness of the GH saddle 
topology at the Euclidean horizon is required by 
the implicit restriction in the path integral 
to the paths that actually label regular states outside a
regular (Lorentzian) horizon. 

\paragraph{What is the correct integration contour?}
As explained above, the path integral for the 
gravitational partition function is not over Euclidean geometries,
and yet the GH saddle is Euclidean.
If the GH saddle point approximation is to find a justification
from first principles, we must presumably start 
by understanding what is the correct integration contour to begin with,
and then attempt to assess whether the GH saddle can be reached 
by an allowed contour deformation. To that end let's get back to basics.

To focus on the nub let's examine the path integral 
computation of the dimension of the Hilbert space, \eqref{TrI},
relevant for the case without any outer boundary (but allowing for a 
horizon boundary).
To warm up, consider a quantum mechanical
system with Hilbert space ${\cal H}$, 
resulting from canonical quantization of a classical
system described by phase space coordinates $(q^i, p_i)$. 
The standard derivation of the path integral representation of
operator matrix elements yields an interesting path integral
representation for the Hilbert space dimension,
\begin{equation}
    Z={\rm Tr}\, I_{\cal H} = \int {\cal D}p_i{\cal D}q^i \, \exp\left(i\oint dt\, p_i\,\dot q^i/\hbar\right)
\end{equation}
where the integral is over all loops in the phase space. 
Note that although it is written in terms of a $t$ parameter above,
the amplitude for a given loop has nothing to do with time dependence; rather, 
it is the exponential of the Poincar\'e invariant,
$\exp\left(i\oint p_i\, dq^i/\hbar\right)$. 

It is straightforward to generalize this 
to field theories. If the theory has a gauge symmetry,  the integral
should be written on the reduced phase space, because otherwise it does 
not compute the trace over
states in the physical Hilbert space.
That is, the gauge constraints should
be imposed and the gauge redundancy eliminated. Faddeev \cite{Faddeev:1969su} worked out the
properties of such an integral and applied it to Yang-Mills  theories,
and Fadeev and Popov \cite{Faddeev:1973zb} later 
applied this construction to general relativity. In both cases, the 
integral can be put in the form
\begin{equation}\label{FP}
    Z={\rm Tr}\, I_{\cal H} = \int {\cal D}p_i{\cal D}q^i \delta(\chi_a)\det\{\chi_b,{\cal C}^c\}
{\cal D}\lambda_a \, \exp\left(i\oint dt(p_i\dot q^i-\lambda^a{\cal C}_a)\right)\,.
\end{equation}
Here the integral is over paths in the unreduced phase space, the indices include spatial position and tensor component labels,
$\chi_a=0$ are gauge fixing conditions, ${\cal C}_a$ are the generators
of gauge transformations, 
$\{\chi_b,{\cal C}^c\}$ is the Poisson bracket of these quantities,
and $\lambda^a$ are ``Lagrange multiplier'' fields whose integral 
results in a delta function $\delta({\cal C}_a)$ which imposes the constraint equations.

There is a pretty wide gulf between the phase space path integral 
\eqref{FP} and the GH saddle. We would like to implement 
a saddle point approximation directly for the integral \eqref{FP},
since that integral hews closest to the Hilbert space calculation,
and it is not entirely clear whether it admits an equivalent 
covariant formulation. 
However, the saddle is presumably stationary, so 
$\dot q^i$ vanishes, and since it is a stationary point of the action,
also the constraints ${\cal C}_a$ vanish. 
Hence the saddle action computed this way 
vanishes, so the calculation fails to yield 
the expected Bekenstein-Hawking result 
for the entropy. Perhaps we
have not properly handled the contribution from the 
horizon boundary. 
In view of this current impasse, 
an alternate strategy is to attempt to 
connect \eqref{FP} to a covariant path integral 
in which the role of the
GH saddle might be more readily identified. 

To recover a covariant path integral from \eqref{FP}
in the case of general relativity, 
the first step is evaluate the integral over the momentum tensor
conjugate to the spatial metric.
In cases where the gauge fixing conditions involve only
the spatial metric \cite{Faddeev:1973zb},
and possibly the trace of the momentum tensor~\cite{Banihashemi:2024aal},
the momentum integrals are
Gaussian, so when carried out they result in replacing the
momentum by its definition in terms of the time derivative
of the spatial metric.
In the exponent this produces the covariant
{\it Lorentzian} action $S_{\rm L}$, with the Lagrange multiplier fields $\lambda^a$ 
playing the role of the lapse and shift which, together with the 
spatial metric, determine the time-time and time-space components of the covariant metric.

One thing this narrative makes clear is that it is essential that 
both signs of the lapse $N := \lambda^0$ must be included on the contour
of integration, because the lapse integral must impose the constraints
in order for the full integral to correspond to a trace on the physical Hilbert space. 
In order for the lapse integral to produce the constraint delta function,
it can traverse any contour that can be deformed to the real line. 
But this statement holds if the $\lambda^0$ integral is carried out 
{\it before} other integrals are evaluated. That, in principle, is the
correct order of integration, because one should be integrating over
the constrained phase space only. However, 
to arrive at a covariant path integral we must carry
out the momentum integrals first. We argued in \cite{Banihashemi:2024aal} that this 
exchange of the order of integration 
is allowed only if the lapse contour is taken to pass below the
origin in the complex lapse plane,
where the integrand of the path integral has an essential singularity.
For example, one can 
replace $N$ by $N - i\epsilon$, with $\epsilon >0$, carry out the momentum
integrals, and then integrate over
real $N$.

The reason for this lapse contour requirement is that 
after gauge-fixing the trace of the momentum,
the Gaussian momentum integrals over the tracefree part
involve the integrand $\sim e^{-iN p^2}$,
and are therefore not convergent unless $N$ has a negative imaginary
part. The shifted $N$ contour also renders the matter momentum
fluctuation integrals convergent, so it appears compatible with the Halliwell-Hartle
requirement \cite{Halliwell:1989dy} 
that the vacuum fluctuations at short distances be controlled 
in the standard way. However it also has the consequence that the 
integration contour is necessarily over metrics that deviate from Lorentzian
by at least a small imaginary part. 
This small imaginary part is actually a welcome feature, not a bug. 
If $Z$ were truly an integral over real metrics weighted by the complex
exponential $e^{iS_{\rm L}}$, then it could not possibly produce the exponential of a 
large positive real number, as needed to recover the Bekenstein-Hawking entropy.

To find an approximation for the dimension of the Hilbert space of states 
of a 3-ball of space, we are thus led to seek a saddle point approximation to the 
almost Lorentzian path integral over metrics on closed loops of time-evolving spatial ball geometries. Since we want to count only the
states that are within the 3-ball and its edge modes,  that system should be 
closed, so there should not be a three dimensional history of the ball boundary
where boundary conditions would have to be imposed.
That is, we want the ball boundary, which is a 2-sphere, to be a fixed point set under the time loop flow. 
This means that the loop of 3-geometries is topologically a 4-sphere, and that the lapse 
function---which determines how much proper time corresponds to a given coordinate time normal to a spatial surface---should vanish at the ball boundary. 
Another reason for the lapse, and also the shift, to vanish (or at least to be fixed)
at the ball boundary, is that the constraints at the boundary relate interior to exterior degrees of freedom, and are hence not relevant to interior observables. Since the
edge modes---would-be gauge degrees of freedom---should be included among the system
degrees of freedom for the interior observer, 
we should not impose the diffeomorphism constraints at the boundary.  
Let's call the ball boundary
the ``horizon'', as that is indeed what it should correspond to in the 
classical saddle solution.

\begin{figure}[h!]
\centering
\includegraphics[width=0.3\textwidth]{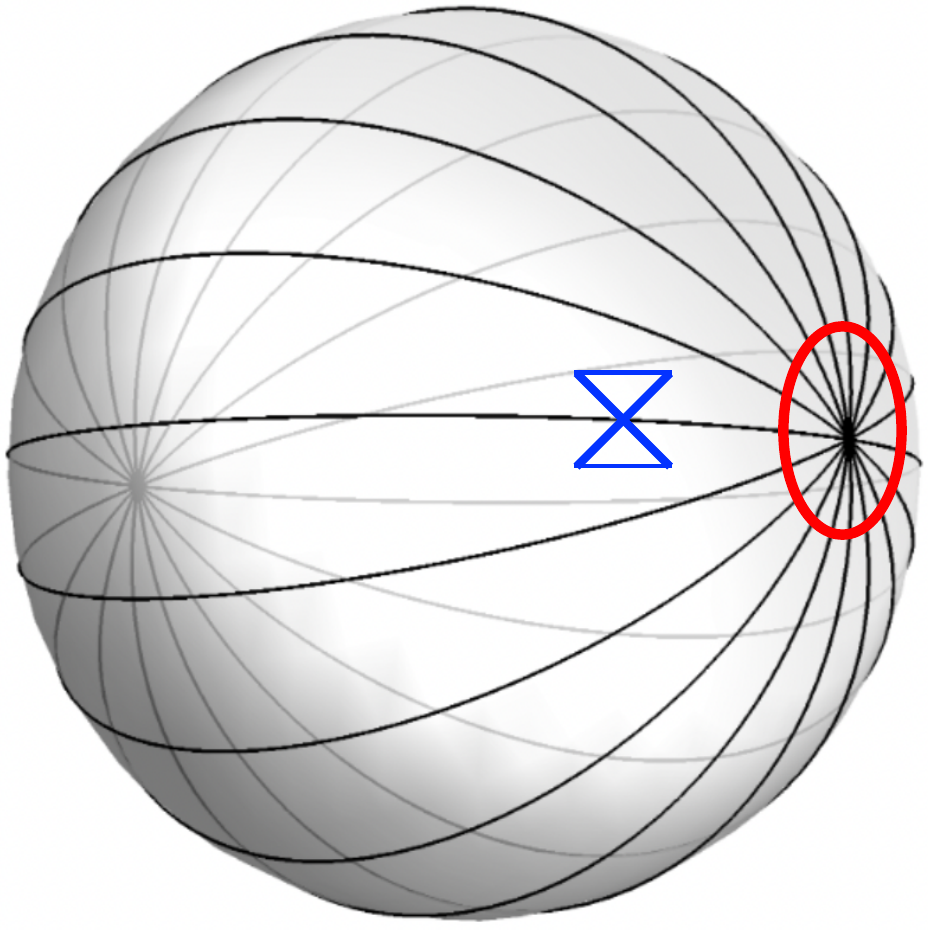}
\caption{Two-dimensional Lorentzian history of a 1-ball, with 1-ball boundary
(the two endpoints) fixed under the time flow. The light cone (blue) shows the
timelike directions. Arbitrarily small closed timelike curves (red) encircle the
endpoints, so no Lorentzian metric can be defined there.
We call such a point a \textit{CTC singularity}.
\label{ctc}}
\end{figure}

The 4-sphere topology agrees with that of the GH saddle, which seems promising.
Worrisome, however, is that no Lorentzian metric exists globally on the 4-sphere. The obstruction
is illustrated in Fig.~\ref{ctc} for the case of 1+1 dimensional spacetime.
It is that there are closed timelike curves
encircling the horizon. It is not that the curvature blows up
at the horizon, but just that there is no smooth extension of the metric to the 
horizon itself, unlike in the Euclidean case. This is similar in some ways to a conical
singularity but, unlike the latter, the irregular tip cannot be realized as a limit 
of smooth metrics, i.e., the tip cannot be ``smoothly rounded off".
We term this irregularity a \textit{CTC singularity}.

But recall that the metric is only {\it almost} Lorentzian: its lapse function should 
have a small negative imaginary part on the original contour. 
The 4-sphere does admit smooth complex metrics, 
so the 
nonexistence of a regular Lorentzian metric on the 4-sphere 
need not indicate a fatal error in the analysis.
The question then is whether there is a complex or Euclidean 
saddle, accessible from the original contour, 
that dominates the integral. 

\begin{figure}[h!]
\centering
\includegraphics[width=0.7\textwidth]{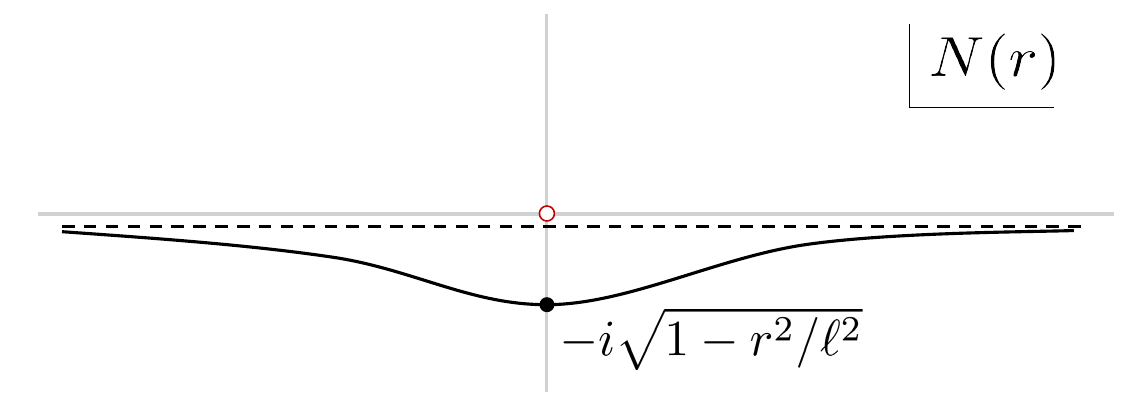}
\caption{Original (dashed) and  deformed  (solid) contour for the lapse integral at each spacetime point.}
\label{contour}
\end{figure}

Note first that, 
apart from the CTC singularity at the horizon, 
a Lorentzian solution on the 4-sphere is provided by simply starting with the de Sitter
metric in a static patch, $ds^2 = -(1-r^2/\ell^2) dt^2 + (1-r^2/\ell^2)^{-1}dr^2 + r^2 d\Omega^2$ (where $\ell= \sqrt{3/\Lambda}$ is the de Sitter radius), 
and identifying two values of the $t$ coordinate. 
As just explained, however, this 
does not extend to a Lorentzian metric at the horizon ($r=\ell$). 
But consider a complex deformation of the de Sitter metric,
\begin{equation}\label{ddS}
   ds^2 = -(1-r^2/\ell^2) \alpha^2  d\tau^2 + (1-r^2/\ell^2)^{-1}dr^2 + r^2 d\Omega^2\,,
\end{equation}
where $\alpha$ is a complex dimensionless constant and 
$\tau$ is a time coordinate of period $2\pi \ell$.
In Lorentzian signature the metric \eqref{ddS} is a solution for any nonzero
real $\alpha$, with a CTC singularity at $r=\ell$, and the smooth Euclidean 
4-sphere corresponds to $\alpha = \pm i$. 
The lapse for the metric \eqref{ddS} in the $\tau$ 
foliation is $N=\alpha \sqrt{1-r^2/\ell^2}$. 
The question is whether the lapse contour at each spacetime point
can be deformed from its original path, which runs just below the real axis,
so as to pass through the Euclidean de Sitter saddle.
Since the contour cannot be deformed across the essential singularity at the origin, we presume it would be the saddle in the lower half plane that dominates the path integral,
as illustrated in Fig.~\ref{contour}. 

\paragraph{Is a complex or Euclidean saddle accessible?}

To know which contour deformations are allowed one needs to know 
the analytic structure of the integrand, i.e., of the complex Einstein-Hilbert action
on a topological 4-sphere. We have not yet analyzed this in the continuum 
setting. However, another way to deal with the CTC singularity 
is to model it with a piecewise flat 
spacetime geometry, using the Regge calculus \cite{Dittrich:2024awu}. 
The contribution to the Regge action from a ``bone'' in $D$ spacetime dimensions
is the $D-2$ dimensional volume of the bone times the deficit angle around the bone,
which is constructed from the dihedral angles of the flat simplices that 
hinge on the bone, which in turn are computed from the squared edge lengths.
The formula for the deficit angle can be analytically continued 
to all simplicial geometries, starting from 
the one that applies for regular Euclidean geometries. Branch points and cuts
arise in this process, when applied to the case of irregular light cone structure. 
This approach was used in Ref.~\cite{Dittrich:2024awu}, where a simplicial microsuperspace 
model of the partition function integral was implemented. It was found there
that, if the original contour is chosen to pass on the side of the branch cut(s) that 
corresponds to adding a negative imaginary part to the lapse, the contour can indeed
be deformed so as to pass through the Euclidean saddle that yields 
(a discrete approximation to) the Bekenstein-Hawking horizon entropy.

\paragraph{Can sense be made of the Lorentzian saddle?}

As explained above, the time-periodic Lorentzian dS metric is a solution
to the Einstein equation, apart from the measure zero horizon set where
the CTC singularity lies. This raises the question of whether, rather than trying to deform the integration contour to pass through a Euclidean saddle, 
we could somehow use that Lorentzian solution to evaluate an approximation to the path integral. We already mentioned our failure thus far to recover the
Bekenstein-Hawking entropy in this manner using the canonical formalism.
To pursue the question using the covariant formalism one
needs to know what action to assign to a spacetime with a CTC singularity. 
One approach, already mentioned above, 
is to discretize the metric using Regge calculus, but 
it should also be possible to work in the continuum.

What needs to be made sense of is the integral of the Ricci
scalar of a two-dimensional Lorentzian geometry---transverse to the horizon---with a CTC singularity. In a related setting, 
Louko and Sorkin \cite{Louko:1995jw}
used a complex regularization of the metric, 
controlled by a parameter which they sent to zero at the end, 
to compute the integral of the Ricci scalar for geometries with a
CTC singularity or other light cone irregularity.
Their result is consistent with a Lorentzian 
version of the Gauss-Bonnet formula that also results from 
analytic continuation of the Regge action.
The same Gauss-Bonnet formula has more recently been implemented in 
\cite{Colin-Ellerin:2020mva} for studies of real-time gravitational
path integrals, and in \cite{Marolf:2022ybi} for gravitational 
thermal partition functions in particular.
Let us describe here how it works.

The Lorentzian Gauss-Bonnet formula  takes the form
\begin{equation}\label{GB}
    \frac12\int_{D^2}  R_2\, dA = {\rm deficit\; angle} = \pm 2\pi i -\oint_{\partial  D^2} k\, ds
\end{equation}
where $D^2$ is a topological disc, $R_2$ is the two-dimensional Ricci scalar,
$dA$ is the area element, 
and $k\, ds$ is the boost angle through which the boundary curve bends.
The sign ambiguity of the $\pm2\pi i$ term is unavoidable.
We discuss below how it is dealt with on physical grounds, 
but first we discuss how it comes about mathematically.

In Euclidean signature the term $\pm 2\pi i$ 
would be instead $2\pi$, the full angle around a regular point. 
In Lorentzian signature, rotations are replaced by boosts, but
boosts do not map a vector across a light ray, so 
one needs to invent a notion of the angle across a light ray
in order to define the full angle around a point. 
This was done in \cite{Sorkin:1975ah} 
(see also \cite{Neiman:2013ap}, \cite{Sorkin:2019llw}, \cite{Dittrich:2024awu} 
for more recent treatments),
where it was seen that there is an imaginary contribution 
$\pm i\pi/2$ for every light ray crossing. 
The ambiguity of the
sign is like an orientation choice that is made once and for all.
When encircling a point in Minkowski
spacetime, the real parts of the boost angles cancel, 
and four light rays are crossed, so 
one obtains the full angle $\pm2\pi i$, which accounts
for the first term on the right hand side of \eqref{GB}.
In Minkowski spacetime the Ricci scalar vanishes, so 
evidently the
second term on the right hand side must cancel the first. It 
does so because there are also imaginary contributions to the 
extrinsic curvature integral when the boundary curve crosses the
light rays. The sign of those contributions is determined
by the detour of the integration contour around a pole in the extrinsic curvature 
at the light rays, so is also ambiguous. The choice of this detour 
must be compatible with the sign
choice made when defining the full imaginary angle around a point.

When the formula \eqref{GB} is applied to a disc with
a CTC singularity at the center, the boundary of the disc crosses
no light rays, so that the $\oint k \, ds$ term 
is real (and equal to the net boost angle). The contribution 
to the Einstein-Hilbert action thus 
picks up an imaginary part, 
\begin{equation}\label{EHi}
    \frac{1}{16\pi G}\int_{M^4} dV_4 \, R_4
    = \pm i A/4G + (\mbox{real})\,,
\end{equation}
where $A$ is area of the submanifold transverse to the disc, i.e., the horizon area.
When this is substituted into the amplitude $\exp({iS_{\rm L}})$, 
in the path integral for $Z$, we obtain 
\begin{equation}\label{Zmp}
    Z\sim \exp(\mp A/4G\hbar)
\end{equation}
which is either totally wrong or wonderfully right!

What considerations determine
which sign in the Lorentzian Gauss-Bonnet formula \eqref{GB} is
the correct one for the physics? 
In \cite{Louko:1995jw} it was shown that the 
ambiguity could be resolved by the requirement that the path integral
for a scalar field on the background of the complex, regularized
metric be convergent, a requirement consistent with that
proposed earlier in \cite{Halliwell:1989dy} on the grounds of
consistency of quantum field theory on the curved spacetime background.
This requirement leads to the minus sign in \eqref{EHi}, 
which is indeed the one required by the Bekenstein-Hawking entropy,
which is in turn intimately related to the entanglement entropy of 
the vacuum of quantum fields. 

While the link between stability of the quantum field theory vacuum and 
the Bekenstein-Hawking entropy is encouraging, 
we should not have to choose the sign 
\textit{a posteriori}---rather, 
it should be determined from first principles.
We have argued above that the lapse contour should pass below the origin
in the complex plane. Perhaps 
this suffices to select the minus sign in 
the action \eqref{EHi} of the Lorentzian CTC singularity, 
but whether that is so remains unclear.

\bmhead{Acknowledgements}
We thank Don Marolf for discussion on the boundedness of the gravitational Euclidean action.
The work of BB was supported by DOE grant DE-SC001010 and the Federico and Elvia Faggin Foundation. The work of TJ was supported in part by NSF grant PHY-2309634.

\bibliography{enigma}

\end{document}